\documentclass[twocolumn]{autart}    

\usepackage{graphicx}          
\usepackage{amsmath}
\usepackage{amsfonts}
\usepackage{listings}
\usepackage{color}
\usepackage{array}
\usepackage{tabu}
\usepackage{cite}
\usepackage{mathrsfs,amssymb}
\usepackage{graphicx}
\graphicspath{ {image22/} }
\usepackage{float}
\usepackage{algorithm}
\usepackage{mathtools}
\usepackage{algpseudocode}
\algdef{SE}[DOWHILE]{Do}{doWhile}{\algorithmicdo}[1]{\algorithmicwhile\ #1}%
\DeclareMathOperator*{\argmin}{arg\,min}
\DeclareMathOperator*{\argmax}{arg\,max}
\DeclareMathOperator*{\card}{card}
\DeclareMathOperator*{\supp}{supp}

\usepackage{ntheorem}

\newtheorem{definition}{Definition}{}
\newtheorem{corollary}{Corollary}{}
\newtheorem{proposition}{Proposition}{}
\newtheorem{theorem}{Theorem}{}
{}
{}
\newtheorem{assumption}{Assumption}{}
{}

\makeatletter

\g@addto@macro\normalsize{%
	
	\setlength\abovedisplayskip{5pt}
	
	\setlength\belowdisplayskip{5pt}
	
}

\makeatother
\begin{document}
\begin{frontmatter}

\title{A Multi-Observer Based Estimation Framework for Nonlinear Systems under Sensor Attacks} 
\thanks{This work was supported by the Australian Research Council under the Discovery Project DP170104099.}

\author[Paestum]{T. Yang}\ead{tianciy@student.unimelb.edu.au},
\author[Paestum]{C. Murguia}\ead{carlos.murguia@unimelb.edu.au},
\author[Paestum]{M. Kuijper}\ead{mkuijper@.unimelb.edu.au},
\author[Paestum]{D. Ne\v{s}i\'{c}}\ead{dnesic@.unimelb.edu.au}

\address[Paestum]{Department of Electrical and Electronics Engineering, The University of Melbourne, Australia}  

\begin{keyword}                           
Nonlinear observers; multi-observer; cyber-physical systems; sensor attacks.             
\end{keyword}                             

\begin{abstract}
	We address the problem of state estimation and attack isolation for general discrete-time nonlinear systems when sensors are corrupted by (potentially unbounded) attack signals. For a large class of nonlinear plants and observers, we provide a general estimation scheme, built around the idea of sensor redundancy and multi-observer, capable of reconstructing the system state in spite of sensor attacks and noise. This scheme has been proposed by others for linear systems/observers and here we propose a unifying framework for a much larger class of nonlinear systems/observers. Using the proposed estimator, we provide an isolation algorithm to pinpoint attacks on sensors during sliding time windows. Simulation results are presented to illustrate the performance of our tools.
\end{abstract}

\end{frontmatter}

\section{Introduction}
 Networked Control Systems (NCSs) have emerged as a technology that combines control, communication, and computation, and offers the necessary flexibility to meet new demands in distributed and large scale systems. Recently, security of NCSs has become a very important issue as wireless communication networks increasingly serve as new access points for adversaries trying to disrupt the system dynamics. Cyber-physical attacks on control systems have caused substantial damage to a number of physical processes. A well-known example is the attack on Maroochy Shire Council's sewage control system in Queensland, Australia. The attacker hacked into the controllers that activate/deactivate valves causing a massive flooding to the surrounding areas. Another more recent incident is the StuxNet virus that targeted Siemens' supervisory control and data acquisition systems which are used in many industrial
processes. These incidents show that strategic mechanisms to identify and deal with attacks on NCSs are needed.\\[1mm]
In \cite{Fawzi2014a,Vamvoudakis2014,Chong2016b,Shoukry2014,Teixeira2012b,Pasqualetti123,Sahand2017, Murguia2016, Carlos_Justin4, Chong2015,Shoukry2016a}, a range of topics related to security of control systems have been discussed. In general, they provide analysis tools for quantifying the performance degradation induced by different classes of attacks and propose reaction and prevention strategies to counter their effect on the system dynamics. Most of the existing work, however, has considered control systems with linear dynamics, although in {many} engineering applications the dynamics of the plants being monitored and controlled is highly nonlinear. There are only a few results addressing the problem of state estimation under attacks for some classes of nonlinear systems. The recent work in \cite{Kim2016a} addresses the problem of sensor attack detection and state estimation for uniformly observable continuous-time nonlinear systems. For a class of power systems under sensor attacks, the authors in \cite{Hu2017} provide an estimator of the system state using compressed sensing techniques. In \cite{Shoukry2016a}, satisfiability modulo theory is used for state estimation for differentially flat systems with corrupted sensors. {In our previous work} \cite{Yang2018a,Yang2018c}, the problem of state estimation and attack isolation for a class of nonlinear systems with \emph{positive-slope nonlinearities} is considered. {We provided} an observer-based estimation/isolation strategy, using a bank of circle-criterion observers, which provides a robust estimate of the system state in spite of sensor attacks and effectively pinpoints attacked sensors.\\[1mm]
The core of our estimation scheme is {based on the work} in \cite{Chong2015}, where the problem of state estimation for \emph{continuous-time LTI systems} is addressed. The authors propose a multi-observer estimator, using a bank of Luenberger observers, which provides a robust estimate of the system state in spite of sensor attacks. In this manuscript, we extend the results in \cite{Chong2015,Yang2018a,Yang2018c} by considering systems with general nonlinear dynamics. We cast the multi-observer estimation scheme in terms of the existence of a bank of (local and practical) nonlinear observers with Input-to-State-Stable (ISS) (with respect to disturbances) {estimator} error dynamics. We consider the setting where the system has $p$ sensors and up to $q<p$ of them are attacked. Following the multi-observer approach given in \cite{Chong2015}, we use a bank of observers to construct an estimator that provides a robust state estimate in the presence of false data injection attacks and noise.\\[1mm]
The main idea behind the multi-observer estimator is the following: Each observer in the bank is driven by a different subset of sensors. Then, for every pair of observers in the bank, the estimator computes the difference between their estimates and selects the observers leading to the smallest difference. If there are attacks {on} some of the sensors, the observers driven by those sensors produce larger differences than the attack-free ones, in general, and thus they are not selected by the estimator. We first consider the noise-free case and show that our estimator converges to the true state of the system in spite of sensor attacks. Next, we consider the case when process disturbances and measurement noise are present. Assuming each observer's error is Input-to-State Stable (ISS) with respect to measurement noise and disturbances in the attack-free case, our estimator provides estimates whose errors satisfy an ISS-like property with respect to disturbances and independent of the attack signals. Compared to the estimation methods given in \cite{Hu2017,Shoukry2016a}, where no system disturbances and noise are considered, our estimation framework can deal with a much larger class of nonlinear systems at the price of having to design multiple observers. Finally, we provide an algorithm for isolating attacked sensors using the proposed estimator and assuming that upper bounds on the system noise are known. The idea behind our isolation algorithm is the following: For each pair of observers, when driven by attack-free sensors, the largest difference between their estimates is proved to be bounded by a threshold that depends on system noise bounds. For every time-step, we select and take the union of all the subsets of sensors such that the corresponding threshold is not crossed; then, the remaining sensors are isolated as attacked ones. {To improve the isolation performance, we carry out the isolation over windows of $N$ time-steps. That is, we select the subset of sensors that is isolated most often in every time window as the attacked ones.} In \cite{Tang2019,Shoukry2015b}, the problem of isolation of attacked sensors for LTI systems is addressed using the \emph{majority-vote} method and \emph{satisfiability modulo theory}, respectively. Compared to those results, our isolation algorithm can be applied to nonlinear and noisy systems.\\[1mm]
The remaining of the paper is organized as follows. Notation is given in Section 2. In Section 3, we present the multi-observer based estimator for the noise-free case. In Section 4, for the case with sensor noise and process disturbances, we prove that the observer-based estimator given in Section 3 provides ISS-like estimates of the system state (with respect to disturbances and noise) that are independent of sensor attacks. An algorithm for attack isolation is given in Section 5. Finally, we give concluding remarks in Section 6.

\section{Notation}
For any vector $v\in\mathbb{R}^{n}$,  we denote {$v^{J}$} the stacking of all $v_{i}$, $i\in J$, $J\subset \left\lbrace 1,\hdots,n\right\rbrace$, $|v|=\sqrt{v^{\top} v}$, and the support set of $v$ as $\supp(v)=\left\lbrace i\in\left\lbrace 1,\hdots,n\right\rbrace |v_{i}\neq0\right\rbrace $. For matrices $C \in\mathbb{R}^{p \times n}$, $C^{\top} = (c_1^{\top},\ldots,c_p^{\top})$, we denote $C^J$ the stacking of all rows $c_{i} \in\mathbb{R}^{1 \times n}$, $i\in J$, $J\subset \left\lbrace 1,\hdots,n\right\rbrace$. For a sequence of vectors $\left\lbrace v(k)\right\rbrace _{k=0}^{\infty}$,  $||v||_{\infty} := \sup_{k\geq 0}|v(k)|$. We say that a sequence $\left\lbrace v(k)\right\rbrace$ belongs to $l_{\infty}$, $\left\lbrace v(k)\right\rbrace \in l_{\infty}$, if $||v||_{\infty}<\infty$. We denote uniformly distributed variables $m$ in the interval $(z_{1},z_{2})$ as $m\sim\mathcal{U}(z_{1},z_{2})$ and normally distributed with mean $\mu$ and variance $\sigma^2$ as $m\sim \mathcal{N}(\mu,\sigma^2)$. A continuous function $\alpha:[0,a)\to[0,\infty)$ is said to belong to class K, if it is strictly increasing and $\alpha(0)=0$, \cite{Khalil:1173048}. Similarity, a continuous function $\beta:[0,a)\times[0,\infty)\to[0,\infty)$ is said to belong to class KL if, for fixed $s$, the mapping $\beta(r,s)$ belongs to class K with respect to $r$ and, for fixed $r$, the mapping $\beta(r,s)$ is decreasing with respect to $s$ and $\beta(r,s)\to 0$ as $s\to\infty$, \cite{Khalil:1173048}.

\section{Multi-Observer Estimator (Noise-free Case)}

A multi-observer based estimator for continuous-time LTI systems has been proposed in \cite{Chong2015}. Similarly, in \cite{Yang2018a}, the authors give an estimator for nonlinear systems with positive-slope nonlinearities. Here, we generalize these results by considering general discrete-time nonlinear systems. Consider the nonlinear system
\begin{equation}\label{s}
\left\{\begin{split}
x^{+}&=f(x,u),\\
y_{i}&=h_{i}(x,u,a_{i}), \hspace{1mm} i\in\left\lbrace 1,\ldots,p\right\rbrace ,
\end{split}\right.
\end{equation}
with state $x\in\mathbb{R}^{n}$, input $u\in\mathbb{R}^{n_{u}}$, $i$-th sensor measurement $y_{i}\in\mathbb{R}$, stacked output $y := (y_1,\ldots,y_p)^{\top} \in \mathbb{R}^{p}$, attack signal $a_{i}\in\mathbb{R}$, {stacked attack vector $a := (a_1,\ldots,a_p)^{\top} \in \mathbb{R}^{p}$ }, and functions $f:\mathbb{R}^{n}\times\mathbb{R}^{n_{u}}\to\mathbb{R}^{n}$ and $h_{i}:\mathbb{R}^{n}\times\mathbb{R}^{n_{u}}\times\mathbb{R}\to\mathbb{R}$. If the $i$-th sensor is not attacked, $a_{i}(k)=0$ for $k\geq 0$; otherwise, sensor $i$ is under attack and $a_{i}(k)$ is arbitrary and possibly unbounded. The \emph{unknown} set of attacked sensors is denoted as $W$, $W\subset\left\lbrace 1,\hdots,p\right\rbrace $. \\[1mm]
\begin{assumption}\label{as1}
The set of attacked sensors does not change over time, i.e.,  $W$ is constant (time-invariant) and $\supp(a(k))\subseteq W$, for all $k\geq 0$.
\end{assumption}
Consider the observer
\begin{equation}\label{o}
\left\{\begin{split}
z_{J}^{+}&=\Gamma_{J}(z_{J},y^{J},u),\\
\hat{x}_{J}&=\eta_{J}(z_{J},y^{J},u),
\end{split}\right.
\end{equation}
where $y^{J}\in\mathbb{R}^{\card(J)}$ denotes the stacking of all $y_{i}$, $i\in J$,\linebreak $J\subset \left\lbrace 1,\hdots,n\right\rbrace$, $z_{J}\in\mathbb{R}^{l_{J}}$ is the observer state, $\hat{x}_{J}\in\mathbb{R}^{n}$ denotes the estimate of the plant state, and $\Gamma_{J}:\mathbb{R}^{l_{J}}\times\mathbb{R}^{\card(J)}\times\mathbb{R}^{n_{u}}\to\mathbb{R}^{l_{J}}$ and $\eta_{J}:\mathbb{R}^{l_{J}}\times\mathbb{R}^{\card(J)}\times\mathbb{R}^{n_{u}}\to\mathbb{R}^{n}$ are some functions.\vspace{2mm}
\begin{definition}\label{def1}\emph{(Local Asymptotic Practical Observer).}
	System \eqref{o} is said to be a local asymptotic practical observer for system \eqref{s} if, for $a^{J}(k)=0$, $k\geq 0$, there exists a set-valued map $\mathcal{D}_{J}(x)\subseteq\mathbb{R}^{l_{J}}$, such that, for any pair of initial conditions $(x(0),z_{J}(0)) \in \mathbb{R}^{n} \times \mathcal{D}_{J}(x(0))$ and $e_{J}(k):=\hat{x}_{J}(k)-x(k)$, there exist KL-function $\beta_{J}(\cdot)$ and $\nu_{J}
\geq0$ satisfying: $|e_{J}(k)|\leq\beta_{J}(|e_{J}(0)|,k)+ \nu_{J}, k\geq 0$.\\[2mm]
\emph{In this manuscript, we assume that observers of form described in Definition \ref{def1} exist and are known for different subsets of sensors $y^J$, $J \subseteq \{1,\ldots,p \}$. Any technique available in literature can be used to construct these observers as long as the corresponding convergence properties satisfy Definition \ref{def1}. Note that all observers guaranteeing global (local) asymptotic convergence satisfy Definition \ref{def1} with $\nu = 0$. In Table 1, we present a list of publications where design methods for nonlinear observers satisfying Definition \ref{def1} are given. We also list the corresponding convergence properties that these observers guarantee. The results in this {paper} apply to all the listed systems/observers.}\vspace{2mm}
\end{definition}
\begin{table}[t]
	\centering
	\begin{tabular}{ |p{3cm}||p{3.5cm}|}
		\hline
		\textbf{\hspace{3.5mm}Convergence} & \textbf{\hspace{4.25mm}References}\\
		\hline
		\hline
		Global exponential &\cite{Ibrir2007b, Moraal1995,Yang2010,Xie1996,Lu2004,Sundaram2016,Yang2018a}\\
		\hline

		Global asymptotic &\cite{Yalcin2015,Xie2016,Zemouche2006,ChaibDraa2017} \\
		\hline
		Local exponential & \cite{Moraal1995,Sundarapandian2002,Sundarapandian2004,Ibrir2009}\\
		\hline
		Local asymptotic  & \cite{GERMANI2009,Califano2003,Ciccarella1995,Ibrir2005,Song1995}\\
		\hline
		Finite-time &\cite{Moraal1995,Kaczorek2017}\\
		\hline
	\end{tabular}
	\vspace{2mm}
	\caption{Systems/observers satisfying Definition \ref{def1} in the literature.}\label{Table1}
\end{table}
\begin{assumption}\label{gg}
	At most $q$ sensors are attacked, i.e.,
	\begin{equation}
	\card(W)\leq q<\frac{p}{2},
	\end{equation}
	where $q$ denotes the largest integer such that for all $J\subset\left\lbrace 1,\ldots,p\right\rbrace$ with $\card(J)\geq p-2q>0$, an observer of the form \eqref{o} exists for any $y^{J}\in\mathbb{R}^{\card(J)}$.\\[2mm]
\emph{Following the ideas in \cite{Chong2015}, we {use a local asymptotic practical observer} for each subset $J\subset\left\lbrace 1,\ldots,p\right\rbrace $ of sensors with $\card(J)=p-q$ and for each subset $S\subset \left\lbrace 1,\ldots,p\right\rbrace $ with $\card(S)=p-2q$. By Assumption \ref{gg}, among the $p$ sensors, there exists at least one subset of sensors $\bar{I}$, \linebreak $\bar{I}\subset\left\lbrace 1,\ldots, p\right\rbrace$,  with $\card(\bar{I})=p-q$ satisfying $y^{\bar{I}}=h^{\bar{I}}(x,u)$, i.e., there is a set $\bar{I}$ of sensors that is attack-free and thus $a^{\bar{I}}(k)=0$ for all $k\geq 0$. Then, in general, the {difference between} estimate $\hat{x}_{\bar{I}}(k)$ and the estimate $\hat{x}_{S}(k)$ given by any subset $S\subset\bar{I}$ with $\card(S)=p-2q$ is smaller than the other subsets $J$ with $\card(J)=p-q$ and $a^{J}(k)\neq 0$. This motivates the following estimation strategy.}
\end{assumption}

\newpage 

For each subset $J\subset\left\lbrace 1,\ldots,p\right\rbrace $ with $\card(J)=p-q$, define $\pi_{J}(k)$ as the largest deviation between the estimates $\hat{x}_{J}(k)$ and $\hat{x}_{S}(k)$ for any $S\subset J$ with $\card(S)=p-2q$:
\begin{equation} \label{54}
\pi_{J}(k):=\max_{S\subset J:\mathbf{\card}(S)=p-2q}|\hat{x}_{J}(k)-\hat{x}_{S}(k)|,
\end{equation}
for all $k\geq 0$, and define the sequence $\sigma(k)$ as
\begin{equation}\label{55}
\sigma(k):=\underset{J\subset\left\lbrace 1,2,\ldots,p\right\rbrace :\card(J)=p-q}{\argmin} \pi_{J}(k). \hspace{2mm}
\end{equation}
Then, as proven below, the estimate indexed by $\sigma(k)$:
\begin{equation}\label{56}
\hat{x}(k)=\hat{x}_{\sigma(k)}(k),
\end{equation}
is an asymptotic attack-free estimate of the system state. The following result {uses the terminology} presented above.
\vspace{3mm}
\begin{theorem}\label{t1} Consider system \eqref{s}, {observer \eqref{o}}, estimator \eqref{54}-\eqref{56}, and the estimation error $e(k)=\hat{x}_{\sigma(k)}(k)-x(k)$. Let Assumption \ref{as1}-\ref{gg} be satisfied; then, there exist a constant $\nu \geq 0$ and a class \emph{KL}-function $\beta(\cdot)$ satisfying:
\begin{equation}
\left\{\begin{split}
|e(k)| &\leq\beta(e_{0},k)+\nu, \label{60}\\
e_{0} &:=\max_{ \tiny{\begin{array}{l} J:\card(J)=p-q \\ S: \card(S)=p-2q \end{array} } } \left\lbrace |e_{J}(0)|, |e_{S}(0)|\right\rbrace,
\end{split}\right.
\end{equation}
for all $k\geq 0$.
\end{theorem}
{We omit the proof of Theorem 1 since we later prove a more general result in Section 4.}

\subsection{Application Examples}

In this subsection, we show the performance of the proposed estimation scheme for two classes of nonlinear systems and observers.\\[2mm]
\textbf{\emph{High Gain Observers:}} Consider the nonlinear system
\begin{equation}
\left\{\begin{split}\label{case1}
x^{+}=&f(x),\\
y=&h(x)+a,
\end{split}\right.
\end{equation}
with state $x\in\mathbb{R}^{n}$, output $y\in\mathbb{R}^{p}$, attack vector $a\in\mathbb{R}^{p}$, and functions $f:\mathbb{R}^{n} \to \mathbb{R}^{n}$ and $h:\mathbb{R}^{n} \to \mathbb{R}^p$.\vspace{2mm}

\begin{assumption}\label{m}
The origin of \eqref{case1} is {locally stable} \emph{\cite{Khalil:1173048}}.\\[2mm]
\emph{Consider the observer
\begin{equation}\label{bbb}
\hat{x}_{J}^{+}=f(\hat{x}_{J})+K_{J}(y^{J}-h(\hat{x}_{J})),
\end{equation}
with state estimate $\hat{x}_J \in\mathbb{R}^{n}$ and observer gain matrix $K_{J}\in\mathbb{R}^{n\times\card(J)}$. The observer gain $K_{J}$ is designed following the results in \cite{Sundarapandian2002}.}\vspace{2mm}
\end{assumption}

\begin{proposition}\label{p1}
	Let Assumption \ref{m} be satisfied and $q$ be the largest integer such that for all $J\subset\left\lbrace 1,\ldots,p\right\rbrace$ with $\card(J)\geq p-2q$ an observer of the form \eqref{bbb} for system \eqref{case1} exists for any $y^{J} \in \mathbb{R}^{\text{\emph{card}}(J)}$. Then, for $a^{J}(k)=0$, $k\geq 0$, there exists a set-valued map $\mathcal{D}_{J}(x)\subsetneq\mathbb{R}^{n}$, such that, for any $(x(0),\hat{x}_{J}(0))\in\mathbb{R}^{n}\times\mathcal{D}_{J}(x(0))$, there are $\lambda_{J}\in(0,1)$ and $c_{J}>0$ satisfying $|e_{J}(k)|\leq c_{J}\lambda_{J}^{k}|e_{J}(0)|$, $k\geq 0$, where $e_{J}=\hat{x}_{J}-x$.\\
	\textbf{Proof:} \emph{Proposition \ref{p1} follows from \cite[Theorem 3]{Sundarapandian2002}.}\\[2mm]
\emph{By Proposition \ref{p1}, system \eqref{case1} with observer \eqref{bbb} satisfy Definition \ref{def1} with $\beta(|e_{J}(0)|,k) = c_{J}\lambda_{J}^{k}|e_{J}(0)|$, $\nu_J=0$, and some set-valued map $\mathcal{D}_{J}(x)$. Hence, we can write the following corollary of Theorem \ref{t1} and Proposition \ref{p1}.}
\end{proposition}
\vspace{2mm}
\begin{corollary}
Consider system \eqref{case1}, observer \eqref{bbb}, the estimator \eqref{54}-\eqref{56}, and the corresponding estimation error $e(k)=\hat{x}_{\sigma(k)}(k)-x(k)$. Let Assumptions \ref{gg} be satisfied; then, there exist $c>0$ and $\lambda\in(0,1)$ satisfying: $|e(k)|\leq c\lambda^{k} e_{0}$, $k\geq 0$, {for $e_{0}$ as defined in \eqref{60}}.\\[2mm]
\emph{\textbf{Example 1:} Consider the following nonlinear system subject to sensor attacks
\begin{equation}
\left\{\begin{split}
x_{1}^{+}=&x_{1}-x_{1}^{3}+x_{2}x_{1}^{2}-x_{2}^{2}x_{1}^{3},\label{mm}\\
x_{2}^{+}=&-x_{2},\\
y_{1}=&2x_{1}+x_{1}^{2},\\
y_{2}=&x_{1}+x_{2}+a_{2},\\
y_{3}=&2x_{1}+x_{2}.
\end{split}\right.
\end{equation}
We have three sensors, i.e., $p=3$. Using the design method given in \cite{Sundarapandian2002}, we have found that observers of the form (\ref{bbb}) exist for each subset $J\subset\left\lbrace 1,2,3\right\rbrace $ with $\card(J)\geq 1$. {By Assumption \ref{gg}, $q=1$, i.e., at most one sensor is attacked.} We let $W=\{2\}$ and design an observer for each $J\subset\left\lbrace 1,2,3\right\rbrace $ with $\card(J)=2$ and each $S\subset\left\lbrace 1,2,3\right\rbrace $ with $\card(S)=1$. Therefore, totally $\binom{3}{2}+\binom{3}{1}=6$ observers are designed. We fix the initial condition of the observers to $\hat{x}(0)=[0,0]^{\top}$, select $(x_{1}(0),x_{2}(0))\in\mathcal{N}(0,1)$, and let $a_{2} \sim\mathcal{U}(-10,10)$. For $k\in[0,49]$, we use \eqref{bbb},(\ref{54})-(\ref{56}) to construct $\hat{x}(k)$. The performance of the estimator is shown in Figure \ref{fig:6e}.\\[1mm]
\begin{figure}[t]
	\centering
\includegraphics[width=0.5\textwidth]{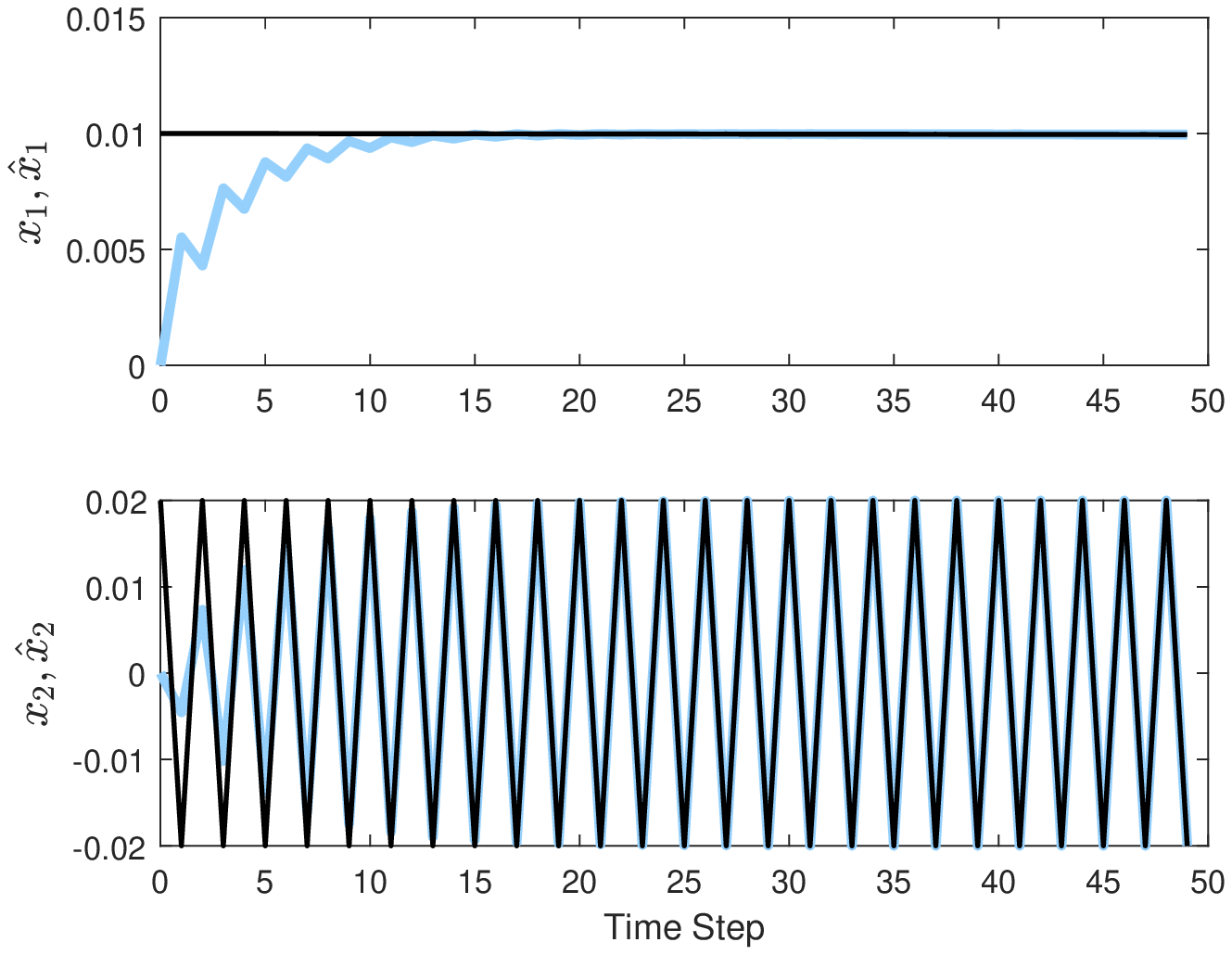}
	\caption{Estimated states $\hat{x}$ converges to the true states $x$ when $a_{2}\sim\mathcal{U}(-10,10)$. Legend: $\hat{x}$ (blue), true states (black).}
	\label{fig:6e}
\end{figure}
\textbf{\emph{Reduced Order Observers:}} Consider the system
\begin{equation}\label{case3s}
\left\{\begin{split}
x^{+}=&Ax+f(x,y),\\
y=&Cx+a,
\end{split}\right.
\end{equation}
with state $x\in\mathbb{R}^{n}$, output $y\in\mathbb{R}^{p}$, attack $a\in\mathbb{R}^{p}$, matrices $A \in \mathbb{R}^{n \times n}$ and $C \in \mathbb{R}^{p \times n}$, and nonlinear function $f:\mathbb{R}^{n} \times \mathbb{R}^{p} \to \mathbb{R}^{n}$.\vspace{2mm}
\begin{assumption}\label{asump6}
$f(x,y)$ is globally Lipschitz in $x$.\\[2mm]
\emph{Consider the partial output vector $y^J = C^Jx + a^J$ and attack $a^J$, with $y^J,a^J \in \mathbb{R}^{\card(J)}$, and the reduced state $\zeta_J = L_Jx \in \mathbb{R}^{n-\text{card}(J)}$, where $L_J \in \mathbb{R}^{(n-\text{card}(J)) \times n}$ is such that $\begin{psmallmatrix} L_J^{\top} & (C^J)^{\top} \end{psmallmatrix}^{\top}$ is nonsingular. Let \[ (N_J,M_J):= \begin{psmallmatrix} L_J \\ C^J \end{psmallmatrix}^{-1};\] then, $x = N_J\zeta_J + M_Jy^J$, and we can write the dynamics of the reduced state $\zeta_J $ as \vspace{2mm}
\begin{equation}\label{reduced}
\zeta_{J}^{+}=A_{L,J}\zeta_{J} + L_{J}\phi_{J}(\zeta_{J},y^{J}) + B_{L,J}y^{J},\\[2mm]
\end{equation}
where $A_{L,J} := L_{J}AN_{J} \in \mathbb{R}^{(n-\card(J)) \times (n-\card(J))}$, $B_{L,J} := L_{J}AM_{J} \in \mathbb{R}^{(n-\card(J)) \times \card(J)}$, and function $\phi_{J}(z_{J},y^{J}) := f(N_{J}z_{J} + M_{J}y^{J},y^{J})$. Consider the reduced order observer\vspace{2mm}
\begin{equation}
	\left\{\begin{split}\label{case3o}
	z_{J}^{+}=&A_{LJ}z_{J}+\phi_{J}(z_{J},y^{J})+B_{LJ}y^{J}\\
              &\hspace{2mm}+K_{J}(y^{J+}-C^{J}\hat{x}_{J}^{+}),\\[1mm]
	\hat{x}_{J}=&N_{J}z_{J}+M_{J}y^{J},
	\end{split}\right.
\end{equation}
with observer state $\hat{z}_{J}\in\mathbb{R}^{n-\card(J)}$, estimated state $\hat{x}_J \in\mathbb{R}^{n}$, and observer matrix $K_{J} \in\mathbb{R}^{(n-\card(J))\times \card(J)}$. We design $K_{J}$ following the results in \cite{Zemouche2006}.}
\end{assumption}
}
\end{corollary}
\vspace{2mm}
\begin{figure}[t]
	\centering
	\includegraphics[width=0.5125\textwidth]{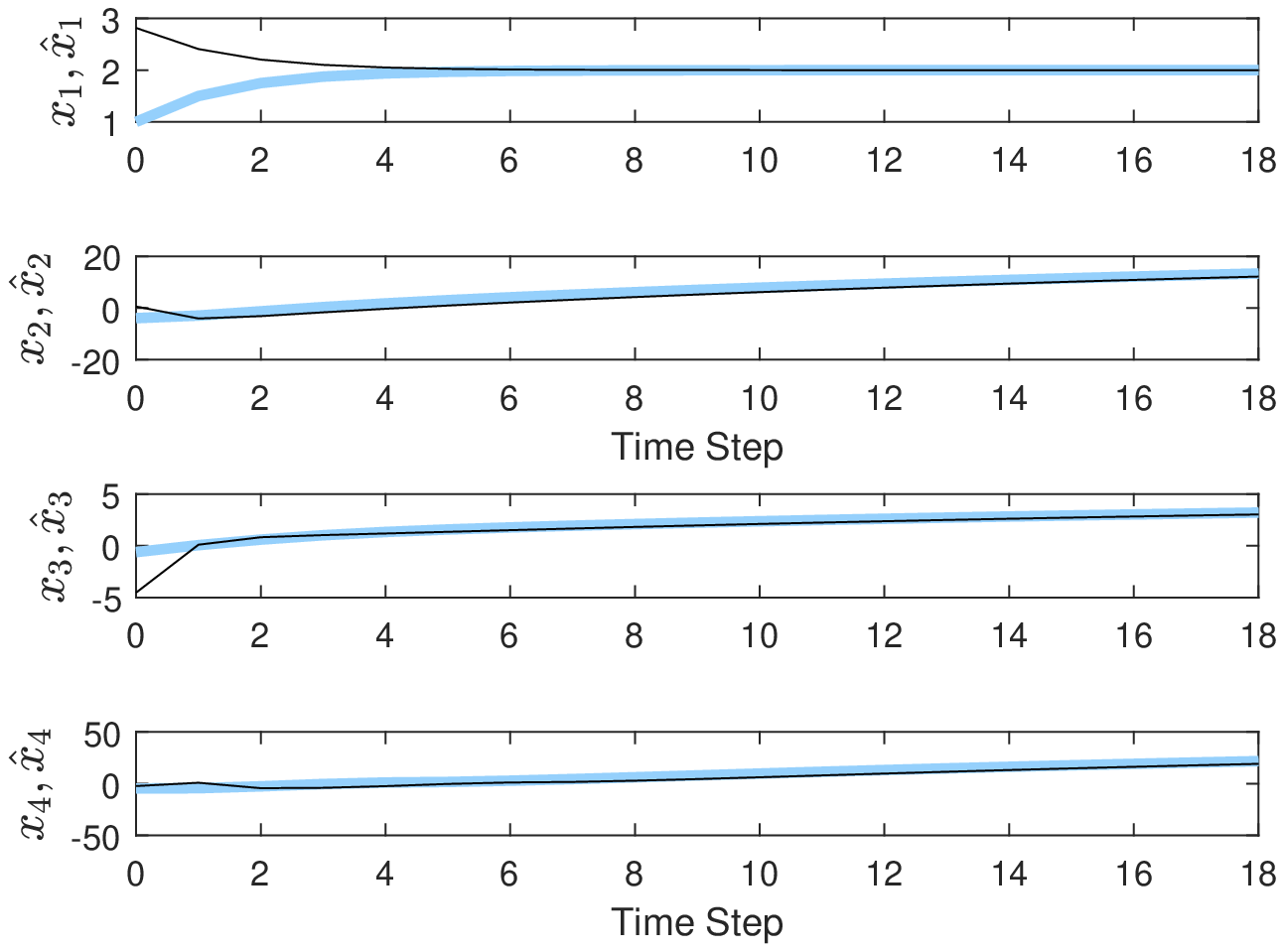}
	\caption{Estimated states $\hat{x}$ converges to the true states $x$ when $a_{2}\sim\mathcal{U}(-10,10)$. Legend: $\hat{x}$ (blue), true states (black)}
	\label{fig:e32}
\end{figure}
\begin{proposition}\label{p3p}
Let Assumption \ref{asump6} be satisfied and $q$ be the largest integer such that for all $J\subset\left\lbrace 1,\ldots,p\right\rbrace$ with $\card(J)\geq p-2q$ an observer of the form \eqref{case3o} for system \eqref{reduced} exists for any $y^{J} \in \mathbb{R}^{\text{\emph{card}}(J)}$. Then, for $a^{J}(k)=0$, $k\geq 0$, and any $(x(0),z_{J}(0))\in\mathbb{R}^{n}\times\mathbb{R}^{l_{J}}$, there exists a KL-function $\beta_{J}(\cdot)$ satisfying: $|e_{J}(k)|\leq\beta_{J}(|e_{J}(0)|,k)$, $k\geq 0$, where $e_{J}=\hat{x}_{J}-x$.\\[2mm]
\textbf{Proof:} \emph{Proposition \ref{p3p} follows from \cite[Theorem 4]{Zemouche2006}.}\\[2mm]
\emph{By Proposition \ref{p3p}, system \eqref{case3s} with observer \eqref{case3o} satisfy Definition \ref{def1} for some KL-function, $\nu_J = 0$, and set-valued map $D_{J}(x)=\mathbb{R}^{n}$. Hence, we can write the following corollary of Theorem \ref{t1} and Proposition \ref{p3p}.}
\end{proposition}
\vspace{2mm}
\begin{corollary}\label{c2}
Consider system \eqref{case3s}, observer \eqref{case3o}, the estimator \eqref{54}-\eqref{56}, and the corresponding estimation error $e(k)=\hat{x}_{\sigma(k)}(k)-x(k)$. Let Assumptions \ref{gg} be satisfied; then, there exists a class KL-function $\beta(\cdot)$ satisfying: $|e(k)|\leq\beta(e_{0},k)$, $k\geq 0$, for $e_{0}$ as defined in \eqref{60}.
\end{corollary}
\textbf{Example 2:} Consider the following nonlinear system under sensor attacks:
	\begin{equation}
	\left\{
	\begin{split}
	x^{+}=&\left[\begingroup 
	\setlength\arraycolsep{2pt}\begin{matrix}
	0.5&0&0&0\\0&0.8&1&0\\0.5&0.1&0.3&0\\0.3&1&0&0.5
	\end{matrix}\endgroup\right] x+\left[ \begin{matrix}
	1\\
	0\\
	0\\
	-1.25\tanh x_{4}-0.6
	\end{matrix}\right], \\
	y=&\left[ \begin{matrix}
	0&1&0&0\\0&0&1&0\\0&0&0&1
	\end{matrix}\right] x+\left[ \begin{matrix}
	0\\a_{2}\\0
	\end{matrix}\right] .
	\end{split}
	\right.
	\end{equation}
Using the design method proposed in \cite{Zemouche2006}, we have found that observers of the form (\ref{case3o}) exist for each subset $J\subset\left\lbrace 1,2,3\right\rbrace $ with $\card(J)\geq 1$ and $p=3$. {By Assumption \ref{gg}, $q=1$, i.e., at most one sensor is attacked.} For randomly selected initial conditions, we attack sensor two, i.e., $W=\left\lbrace 2\right\rbrace $, and let $a_{2}\sim\mathcal{U}(-10,10)$. We use \eqref{case3o}, (\ref{54})-(\ref{56}) to reconstruct $x(k)$. The performance of the estimator is shown in Figure \ref{fig:e32}.

\section{Robust Multi-Observer Based Estimator}

The tools given in this section, generalize the results in \cite{Chong2015,Yang2018a} by considering systems with general nonlinear dynamics, disturbances, and noise. Consider the system
\begin{equation}
\begin{split}\label{g}
{x}^{+}&=F({x},u,d),\\
{y}_{i}&=g_{i}({x},u,m_{i},a_{i}), i\in\left\lbrace 1,\ldots,p\right\rbrace ,
\end{split}
\end{equation}
with state $x\in\mathbb{R}^{n}$, input $u\in\mathbb{R}^{n_{u}}$, disturbance $d\in\mathbb{R}^{s}$, $\left\lbrace d(k)\right\rbrace \in l_{\infty}$, $i$-th sensor measurement $y_{i}\in\mathbb{R}$, stacked measurements $y := (y_1,\ldots,y_p)^{\top} \in \mathbb{R}^{p}$, attack signal $a_{i}\in\mathbb{R}$, measurement noise $m_{i}\in\mathbb{R}$, $\left\lbrace m_{i}(k)\right\rbrace \in l_{\infty}$, and nonlinear functions $F:\mathbb{R}^{n}\times\mathbb{R}^{n_{u}}\times\mathbb{R}^{s}\to\mathbb{R}^{n}$ and $g_{i}:\mathbb{R}^{n}\times\mathbb{R}^{n_{u}}\times\mathbb{R}\times\mathbb{R}\to\mathbb{R}$.\\[1mm]
Consider the observer
\begin{equation}\label{o3}
\left\{\begin{split}
{z}_{J}^{+}&={\Gamma}_{J}({z}_{J},{y}^{J},u),\\
\hat{{x}}_{J}&={\eta}_{J}({z}_{J},{y}^{J},u),
\end{split}\right.
\end{equation}
 where ${z}_{J}\in\mathbb{R}^{l_{J}}$ is the observer state, $\hat{{x}}_{J}\in\mathbb{R}^{n}$ denotes the state estimate, and ${\Gamma}_{J}:\mathbb{R}^{l_{J}}\times\mathbb{R}^{\card(J)}\times\mathbb{R}^{n_{u}}\to\mathbb{R}^{l_{J}}$ and ${\eta}_{J}:\mathbb{R}^{l_{J}}\times\mathbb{R}^{\card(J)}\times\mathbb{R}^{n_{u}}\to\mathbb{R}^{n}$ are some functions.\\[.1mm]
\begin{definition}\label{def2}\emph{(Local ISS Practical Observer).}
System \eqref{o3} is said to be a local asymptotic practical observer for system \eqref{g} if, for $a^{J}(k)=0$, $k\geq 0$, there exists a set-valued map $\mathcal{D}_{J}(x)\subseteq\mathbb{R}^{l_{J}}$, such that for any pair of initial conditions $(x(0),z_{J}(0)) \in \mathbb{R}^{n} \times \mathcal{D}_{J}(x(0))$ and \linebreak $e_{J}=\hat{x}_{J}-x$, there exist a \emph{KL}-function $\beta_{J}(\cdot)$, \emph{K}-functions $\gamma_{1,J}(\cdot)$ and $\gamma_{2,J}(\cdot)$, and constant $\nu_{J}\
	\geq0$ satisfying:
	\begin{equation}\label{xx}
	\begin{split}
	|e_{J}(k)|\leq &\beta_{J}(|{e}_{J}(0)|,k)+\gamma_{1,J}(||m^{J}||_{\infty})\\[1mm] &\hspace{5mm}+\gamma_{2,J}(||d||_{\infty})+\nu_{J},\hspace{1mm} k\geq 0.
	\end{split}
	\end{equation}
\emph{We assume that observers of form given in Definition \ref{def2}\linebreak exist and are known for different subsets of sensors $y^J$, $J \subseteq \{1,\ldots,p \}$. In Table 2, we present a list of references where design methods for nonlinear observers satisfying Definition \ref{def2} can be found. All these observers can be used to construct the proposed estimator.
}
\vspace{2mm}
\end{definition}
\begin{assumption}\label{gg2}
At most $q$ sensors are attacked, i.e.,
\begin{equation}
\card(W)\leq q<\frac{p}{2},
\end{equation}
where $q$ denotes the largest integer such that for all $J\subset\left\lbrace 1,\ldots,p\right\rbrace$ with $\card(J)\geq p-2q>0$, an observer of the form \eqref{o3} exists for any $y^{J}\in\mathbb{R}^{\card(J)}$.\vspace{2mm}
\end{assumption}
\begin{theorem} Consider system \eqref{g}, observer \eqref{o3}, estimator \eqref{54}-\eqref{56}, and the  estimation error ${e}(k)=\hat{{x}}_{\sigma(k)}(k)-{x}(k)$. Let Assumptions \ref{gg2} be satisfied; then, there exist a class \emph{KL}-function $\beta(\cdot)$, class \emph{K}-functions $\gamma_{1}(\cdot)$ and $\gamma_{2}(\cdot)$, and a constant $\nu \geq 0$ satisfying: \label{bb}
\begin{equation}
\left\{\begin{split}
|e(k)| &\leq\beta(e_{0},k)+\gamma_{1}(||m||_{\infty})+\gamma_{2}(||d||_{\infty})+\nu, \label{61}\\
e_{0} &:=\max_{ \tiny{\begin{array}{l} J:\card(J)=p-q \\ S: \card(S)=p-2q \end{array} } } \left\lbrace |e_{J}(0)|, |e_{S}(0)|\right\rbrace.
\end{split}\right.
\end{equation}
for all $k\geq 0$ and $\left\lbrace m(k)\right\rbrace,\left\lbrace d(k)\right\rbrace \in\l_{\infty}$.
\end{theorem}
\begin{table}[t]
	\centering
	\begin{tabular}{ |p{3cm}||p{3.0cm}|}
		\hline
		\textbf{\hspace{3.5mm}Convergence} & \textbf{\hspace{4.25mm}References}\\
		\hline
		\hline
		Global exponential &
		\cite{Xie1996,Sundaram2016,Lu2004,Yang2018a}\\
		\hline
		Global asymptotic &
		\cite{Abbaszadeh2008}\\
		\hline
	\end{tabular}
	\vspace{2mm}
	\caption{Systems/observers satisfying Definition \ref{def2} in the literature.}\label{Table2}
\end{table}
\emph{\textbf{Proof:}} Under Assumption \ref{gg2},  there exist at least one subset $\bar{I}$ with $\card(\bar{I})=p-q$ and $a^{\bar{I}}(k)=0$ for all $k\geq 0$. Then, {by definition \ref{def2}}, there exist a KL-function $\beta_{\bar{I}}(\cdot)$, class K-functions $\gamma_{1,\bar{I}}(\cdot)$ and $\gamma_{2,\bar{I}}(\cdot)$, and $\nu_{\bar{I}}\geq 0$ such that
\begin{equation}
|{e}_{\bar{I}}(k)|\leq\beta_{\bar{I}}(e_{0},k)+\gamma_{1,\bar{I}}(||m^{\bar{I}}||_{\infty})+\gamma_{2,\bar{I}}(||d||_{\infty})+\nu_{\bar{I}},
\end{equation}
for all $k\geq0$. For all $S\subset\bar{I}$ with $\card(S)=p-2q$, there exist a KL-function $\beta_{S}(\cdot)$, class K-functions $\gamma_{1,S}(\cdot)$ and $\gamma_{2,S}(\cdot)$, and $\nu_{S}\geq 0$ such that
\begin{equation}
|{e}_{S}(k)|\leq\beta_{S}(e_{0},k)+\gamma_{1,S}(||m^{S}||_{\infty})+\gamma_{2,S}(||d||_{\infty})+\nu_{S},
\end{equation}
for all $k\geq0$, which yields
\begin{equation}
\begin{split}
\pi_{\bar{I}}(k)=&\underset{S\subset\bar{I}}{\max}|\hat{x}_{\bar{I}}(k)-\hat{x}_{S}(k)|\\
=&\underset{S\subset\bar{I}}{\max}|\hat{{x}}_{\bar{I}}(k)-{x}(k)+x(k)-\hat{x}_{S}(k)|\\
\leq&  |e_{\bar{I}}(k)|+\underset{S\subset\bar{I}}{\max}|e_{S}(k)|\\
\leq& 2(\beta'(e_{0},k)+\gamma'_{1}(||m^{\bar{I}}||_{\infty})+\gamma'_{2}(||d||_{\infty})
+\nu'),\label{e}
\end{split}
\end{equation}
for all $k\geq 0$, where \[\gamma'_{1}(||m^{\bar{I}}||_{\infty})=\underset{S\subset\bar{I}}{\max}\left\lbrace \gamma_{1,\bar{I}}(||m^{\bar{I}}||_{\infty}),\gamma_{1,S}(||m^{\bar{I}}||_{\infty})\right\rbrace,\] \[\gamma'_{2}(||d||_{\infty})=\underset{S\subset\bar{I}}{\max}\left\lbrace \gamma_{2,\bar{I}}(||d||_{\infty}),\gamma_{2,S}(||d||_{\infty})\right\rbrace.\]
Under Assumption \ref{gg2}, for each subset $J\subset\left\lbrace 1,\ldots,p\right\rbrace $ with $\card(J)=p-q$, there exists $\bar{S}\subset J$ with $\card(\bar{S})=p-2q$ such that $a^{\bar{S}}(k)=0$ for all $k\geq 0$, and there exist a KL-function $\beta_{\bar{S}}(\cdot)$, class K-functions $\gamma_{1,\bar{S}}(\cdot)$ and $\gamma_{2,\bar{S}}(\cdot)$, and $\nu_{\bar{S}}\geq 0$ such that
\begin{equation}\label{f}
|e_{\bar{S}(k)}|\leq\beta_{\bar{S}}(e_{0},k)+\gamma_{1,\bar{S}}(||m^{\bar{S}}||_{\infty})+\gamma_{2,\bar{S}}(||d||_{\infty})+\nu_{\bar{S}},
\end{equation}
for all $k\geq 0$.  From (\ref{54}), by construction \[
\begin{split}
\pi_{\sigma(k)}(k)=&\underset{S\supset \sigma(k):\card(S)=2q}{\max}|\hat{x}_{\sigma(k)}(k)-\hat{x}_{S}(k)|\\\geq&|\hat{x}_{\sigma(k)}(k)-\hat{x}_{\bar{S}}(k)|,
\end{split}
\] using the above lower bound on $\pi_{\sigma(k)}(k)$ and the triangle inequality, we have that
\begin{equation}
\begin{split}
|e_{\sigma(k)}(k)|=&|\hat{x}_{\sigma(k)}(k)-x(k)|\\
=&|\hat{x}_{\sigma(k)}(k)-\hat{x}_{\bar{S}}(k)+\hat{x}_{\bar{S}}(k)-x(k)|\\
\leq&|\hat{x}_{\sigma(k)}(k)-\hat{x}_{\bar{S}}(k)|+|e_{\bar{S}}(k)|\\
\leq&\pi_{\sigma(k)}(k)+|e_{\bar{S}}(k)|\\
\leq&\pi_{\bar{I}}(k)+|e_{\bar{S}}(k)|,
\end{split}
\end{equation}
for all $k\geq 0$. Hence, from (\ref{e}) and (\ref{f}), we have
\begin{equation}
|e_{\sigma(k)}|\leq 3(\beta_{1}(e_{0},k)+\gamma_{1,1}(||m||_{\infty})+\gamma_{2,1}(||d||_{\infty})+\nu_{1}),\label{hh}
\end{equation}
for all $k\geq 0$, where
\[\gamma_{1,1}(||m||_{\infty})=\max\left\lbrace \gamma'_{1}(||m||_{\infty}),\gamma_{1,\bar{S}}(||m||_{\infty})\right\rbrace,\]
\[\gamma_{2,1}(||d||_{\infty})=\max\left\lbrace \gamma'_{1}(||d||_{\infty}),\gamma
_{1,\bar{S}}(||d||_{\infty})\right\rbrace.\]
Inequality \eqref{hh} is of the form (\ref{61}) with KL-function $\beta(e_{0},k)=3\beta_{1}(e_{0},k)$, nonnegative constant $\nu=3\nu_{1}$, and K-functions  $\gamma_{1}(||m||_{\infty})=3\gamma_{1,1}(||m||_{\infty})$, and $\gamma_{2}(||d||_{\infty})=3\gamma_{2,1}(||d||_{\infty})$. \hfill$\blacksquare$

\subsection{Application Example}
\begin{figure}[t]
		\centering
		\includegraphics[width=0.5\textwidth]{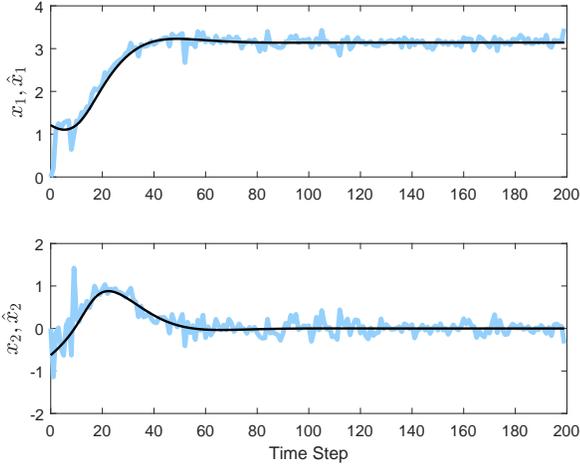}
		\caption{Estimated states $\hat{x}$ converges to a neighbourhood of the true states $x$ when $(a_{2},a_{5})\sim\mathcal{U}(-10,10)$. Legend: $\hat{x}$ (blue), true states (black)}
		\label{fig:10e}
	\end{figure}

The following class of systems has been included in our preliminary work [36].\\[2mm]
\textbf{\emph{Circle-Criterion Observers}:}
Consider the system
\begin{equation}\label{case4}
\left\{\begin{split}
x^{+}=&Ax+Gf(Hx)+\rho(u,y),\\
y=&Cx+a+m,
\end{split}\right.
\end{equation}
with state $x\in\mathbb{R}^{n}$, control $u\in\mathbb{R}^{n_{u}}$, output $y\in\mathbb{R}^{p}$, measurement noise $m\in\mathbb{R}^{p}$, $\left\lbrace m(k)\right\rbrace \in l_{\infty}$, and matrices $G\in\mathbb{R}^{n\times r}$ and $H\in\mathbb{R}^{r\times n}$. The term $\rho(u,y)$ is a known arbitrary real-valued vector that depends on the system inputs and outputs. The state-dependent nonlinearity $f(Hx)$ is an $r$-dimensional vector which each entry is a function of a linear combination of the states:
\begin{equation}\label{230}
f_{i}=f_{i}\left( \sum_{j=1}^{n}H_{ij}x_{j}\right) ,\quad i=1,\ldots,r
\end{equation}
where $H_{ij}$ are the entries of matrix $H$.
\vspace{2mm}
\begin{assumption}\label{1000} For any $i\in\left\lbrace 1,\ldots,r\right\rbrace $:
	\begin{equation}
	\frac{f_{i}(v_{i})-f_{i}(w_{i})}{v_{i}-w_{i}}\geq 0,\hspace{1mm} \forall \hspace{1mm} v_{i},w_{i}\in\mathbb{R},\hspace{1mm} v_{i}\neq w_{i}.\\[3mm]\label{14}
	\end{equation}
\emph{Consider the circle-criterion observer
\begin{equation}
\begin{split}
\hat{x}_{J}^{+}=&A\hat{x}_{J}+Gf\big(H\hat{x}_{J}+K_{J}(C^{J}\hat{x}_{J}-y^{J})\big)\\
&+L_{J}(C^{J}\hat{x}_{J}-y^{J})+\rho(u,y),\label{88}
\end{split}
\end{equation}
with estimated state $\hat{x}_{J}\in\mathbb{R}^{n}$ and observer gain matrices $K_{J}\in\mathbb{R}^{r\times\card(J)}$ and $L_{J}\in\mathbb{R}^{n\times\card(J)}$. Matrices $K_{J}$ and $L_{J}$ are designed following the results in \cite{Yang2018a}.}
\end{assumption}
\vspace{3mm}
\begin{proposition}\label{p5}
Let Assumption \ref{1000} be satisfied, and $q$ be the largest integer such that for all $J\subset\left\lbrace 1,\ldots,p\right\rbrace $ with $\card(J)\geq p-2q>0$ an observer of the form \eqref{88} for system \eqref{case4} exists for any $y^{J}\in\mathbb{R}^{\card(J)}$. Then, for $a^{J}(k)=0$, $k\geq 0$, and any $(x(0),\hat{x}_{J}(0))\in\mathbb{R}^{n}\times\mathbb{R}^{n}$, there exist $c_{J}>0$, $\lambda_{J}\in(0,1)$, and $\gamma_{1,J}>0$ satisfying:
	$|e_{J}(k)|\leq c_{J}\lambda_{J}^{k}|{e}_{J}(0)|+\gamma_{1,J}||m||_{\infty}$, $k\geq 0$, $\left\lbrace m(k)\right\rbrace \in l_{\infty}$, where $e_{J}=\hat{x}_{J}-x$.\\[3mm]
\textbf{Proof:} \emph{Proposition \ref{p5} follows from \cite[Theorem 1]{Yang2018a}.\\[3mm]
By Proposition \ref{p5}, system \eqref{case4} with observer \eqref{88} satisfy Definition \ref{def2} with $\beta(|e_{J}(0)|,k) = c_{J}\lambda_{J}^{k}|e_{J}(0)|$, constant $d=0$, linear function $\gamma_{1,J}$, $\nu_J = 0$, and set-valued map $D_{J}(x)=\mathbb{R}^{n}$. Hence, we can write the following corollary of Theorem \ref{bb} and Proposition \ref{p5}.}
\end{proposition}
\vspace{3mm}
\begin{corollary}
Consider system \eqref{case4}, observer \eqref{88}, the estimator \eqref{54}-\eqref{56}, and the corresponding estimation error $e(k)=\hat{x}(k)_{\sigma(k)}-x(k)$. Let Assumptions \ref{gg2} be satisfied; then, there exist $c>0$, $\lambda\in(0,1)$, $\gamma_{1}>0$ satisfying: $
	|{e}(k)|\leq c\lambda^{k} e_{0}+\gamma_{1}||m||_{\infty}$, $k\geq 0$, $\left\lbrace m(k)\right\rbrace \in l_{\infty}$, for $e_{0}$ as defined in \eqref{61}.
\end{corollary}
\textbf{Example 3:} Consider the following system subject to sensor noise and attacks
	\begin{equation}
	\left\{\begin{split}
	x^{+}=&\left[ \begin{matrix}
	1&0.1\\
	0&1
	\end{matrix}\right]x+\left[ \begin{matrix}
	0.05 \sin (x_{1}+x_{2})\\
	0.1\sin (x_{1}+x_{2})
	\end{matrix}\right],
	\label{mmm} \\
	y=&\left[ \begin{matrix}
	3&3&6&1.2&1.5\\
	0.3&0.6&0.9&12&15\\
	\end{matrix}\right]^{\top} x+m+a,
	\end{split}\right.
	\end{equation}
with $m_{i}\sim\mathcal{U}(-0.1,0.1),i\in\left\lbrace 1,\ldots,5\right\rbrace $. Using the design method proposed in \cite{Yang2018a}, we have found that observers of the form (\ref{88}) exist for each subset $J\subset\left\lbrace 1,2,3,4,5\right\rbrace $ with $\card(J)\geq 1$ and $p=5$. By Assumption \ref{gg2}, $q=2$, i.e., at most two sensors are attacked. We design an observer for each $J\subset\left\lbrace 1,2,3,4,5\right\rbrace $ with $\card(J)=3$ and each $S\subset\left\lbrace 1,2,3,4,5\right\rbrace $ with $\card(S)=1$. Therefore, totally  $\binom{5}{3}+\binom{5}{1}=15$ observers are designed. We attack sensors two and five, i.e., $W=\left\lbrace 2,5\right\rbrace $, and let $(a_{2},a_{5})\sim\mathcal{U}(-10,10)$. For $k\in[0,199]$, we use \eqref{88},(\ref{54})-(\ref{56}) to construct $\hat{x}(k)$. The performance of the estimator is shown in Figure \ref{fig:10e}.

\section{Isolation of Attacked Sensors}
\begin{figure}[t]
		\centering
		\includegraphics[width=0.5\textwidth]{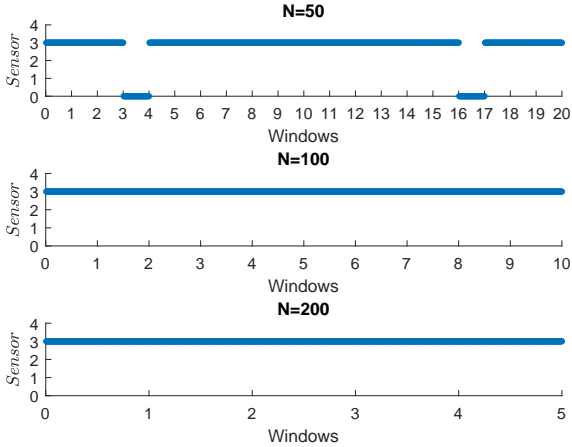}
		\caption{Attack isolation, $a_{3}\sim\mathcal{U}(-2,2)$.}
		\label{fig:1}
		\centering
	\end{figure}

{Using the proposed estimation scheme, in our previous work \cite{Yang2018c}, for a class of nonlinear systems with positive-slope nonlinearities, we have provided an algorithm for isolating sensor attacks. Here, we generalize this algorithm to deal with the larger class of systems \eqref{g}.} Consider system (\ref{g}) and let $q$ be the largest integer such that an observer of the form (\ref{o3}) satisfying Definition \ref{def2} exists for each subset $J\subset\left\lbrace 1,\ldots,p\right\rbrace $ with $\card(J)\geq p-2q$.\\[2mm]
\begin{assumption}\label{ld}
	Bounds on the process disturbance $d$ and the sensor noise $m$ are known, i.e.,
	\begin{equation}
	||d||_{\infty}=\bar{d}, \hspace{1mm}||m||_{\infty}=\bar{m},
	\end{equation}
where $\bar{d}\geq 0$ and $\bar{m}\geq 0$ are known constants.
\end{assumption}
To perform the isolation, we construct an observer satisfying Definition \ref{def2} for each subset $J\subset\left\lbrace 1,\ldots,p\right\rbrace $ of sensors with $\card(J)=p-q$ and each subset $S\subset\left\lbrace 1,\ldots,p\right\rbrace $ with $\card(S)=p-2q$. Hence, by Definition \ref{def2}, for $a^{S}(k)=0$, $k\geq 0$, there exist a KL-function, $\beta_{S}(\cdot)$, K-functions, $\gamma_{1,S}(\cdot)$ and $\gamma_{2,S}(\cdot)$, and $\nu_{S}\geq 0$ satisfying:\vspace{1mm}
\begin{equation}\label{i1}
|e_{S}(k)|\leq \beta_{S}(|e(0)|,k)+\gamma_{1,S}(\bar{m})+\gamma_{2,S}(\bar{d})+\nu_{S},\\[1mm]
\end{equation}
for all $k\geq0$. Note that, there always exist a $k^{*}_{S}$ such that
$
\beta_{S}(|e(0)|,k)\leq\epsilon,
$
for any $\epsilon>0$ and $k\geq k_{S}^{*}$. Then,
\begin{equation}\label{xxxx}
|e_{S}(k)|\leq \epsilon+\gamma_{1,S}(\bar{m})+\gamma_{2,S}(\bar{d})+\nu_{S},
\end{equation}
for all $k\geq k_{S}^{*}$.
Define $\bar{k}^{*}:=\max_{J,S}\left\lbrace k_{J}^{*},k_{S}^{*}\right\rbrace$. By Assumption \ref{gg2}, there are at most $q$ sensors under attack; then, we know there exists at least one $\bar{I}\subset\left\lbrace 1,\ldots,p\right\rbrace $ with $\card(\bar{I})=p-q$ such that $a^{\bar{I}}(k)=0, k\geq 0$, and \vspace{1mm}
\begin{equation}\label{i2}
|e_{\bar{I}}(k)|\leq\epsilon+\gamma_{1,\bar{I}}(\bar{m})+\gamma_{2,\bar{I}}(\bar{d})+\nu_{\bar{I}},\\[1mm]
\end{equation}
for all $k\geq k_{\bar{I}}^{*}$. Then, we have
\begin{equation}
\begin{split}
\pi_{\bar{I}}(k):=&\underset{S\subset\bar{I}}{\max}|\hat{x}_{\bar{I}}(k)-\hat{x}_{S}(k)|\\
=&\underset{S\subset\bar{I}}{\max}|\hat{x}_{\bar{I}}(k)-x(k)+x(k)-\hat{x}_{S}(k)|\\
\leq&  |e_{\bar{I}}(k)|+\underset{S\subset\bar{I}}{\max}|e_{S}(k)|.
\end{split}
\end{equation}
From (\ref{xxxx}) and (\ref{i2}), we obtain
\[
\pi_{\bar{I}}(k)\leq 2(\epsilon+\gamma_{1,\bar{I}}'(\bar{m})+\gamma_{2,\bar{I}}'(\bar{d})+\nu_{\bar{I}}'),
\]
for all $k\geq\bar{k}^{*}$, where
\[
\gamma_{1,\bar{I}}'(\bar{m}):=\underset{S\subset\bar{I}:\card(S)=p-2q}{\max}\left\lbrace \gamma_{1,\bar{I}}(\bar{m}), \gamma_{1,S}(\bar{m})\right\rbrace,
\]
and
\[
\gamma_{2,\bar{I}}'(\bar{d}):=\underset{S\subset\bar{I}:\card(S)=p-2q}{\max}\left\lbrace \gamma_{2,\bar{I}}(\bar{d}), \gamma_{2,S}(\bar{d})\right\rbrace.
\]
However, if the subset $J$ of sensors is under attack at time $k$, i.e., $a^{J}(k)\neq 0$, then $\hat{x}_{J}(k)$ and $\hat{x}_{S}(k)$ in $\pi_{J}(k)$ are more inconsistent and produce larger $\pi_{J}(k)$. Define \vspace{1mm}
\begin{equation}\label{ggs}
\bar{\pi}_{J} :=2(\epsilon+\gamma_{1J}'(\bar{m})+\gamma_{2J}'(\bar{d})+\nu_{J}'),\\[1mm]
\end{equation}
for each $J\subset\left\lbrace 1,\ldots,p\right\rbrace $ with $\card(J)=p-q$, where
\[
\gamma_{1,J}'(\bar{m}):=\underset{S\subset J:\card(S)=p-2q}{\max}\left\lbrace \gamma_{1,J}(\bar{m}), \gamma_{1,S}(\bar{m})\right\rbrace,
\]
and
\[
\gamma_{2,J}'(\bar{d}):=\underset{S\subset J:\card(S)=p-2q}{\max}\left\lbrace \gamma_{2,J}(\bar{d}), \gamma_{2,S}(\bar{d})\right\rbrace;
\]
then, $\bar{\pi}_{J}$ can be used as a threshold to isolate attacked sensors. For all $k\geq\bar{k}^{*}$, we select from all the subsets $J\subset\left\lbrace 1,\ldots,p\right\rbrace $ with $\card(J)=p-q$, the ones that satisfy
\begin{equation}\label{777}
\pi_{J}(k)\leq\bar{\pi}_{J}.
\end{equation}
Denote as $\bar{W}(k)$ \emph{the set of sensors that we regard as attack-free at time $k$}. We construct $\bar{W}(k)$ as the union of all subsets $J$ satisfying (\ref{777}):
\begin{equation}\label{ccc}
\bar{W}(k):=\underset{J\subset\left\lbrace 1,\ldots,p\right\rbrace :\card(J)=p-q,\pi_{J}(k)\leq\bar{\pi}_{J}}{\bigcup} J.
\end{equation}
Thus, the set $\left\lbrace 1,\ldots,p\right\rbrace \setminus\bar{W}(k)$ is isolated as the set of attacked sensors at time $k$. Note, however, that, for small persistent attacks, it is still possible that for some $k\geq\bar{k}^{*}$ and some $J\subset \left\lbrace 1,\ldots, p\right\rbrace $ with $\card(J)=p-q$, $a^{J}(k)\neq 0$ but (\ref{777}) still holds. This implies that $J\subset\bar{W}(k)$ even if $a^{J}(k)\neq 0$ and would result in wrong isolation at time $k$. To improve the isolation performance, we carry out the isolation over windows of $N$ time-steps, $N \in \mathbb{N}$. That is, for each $k \in [\bar{k}^{*}+(i-1)N,\bar{k}^{*}+iN]$, $i \in \mathbb{N}$, we compute and collect $\bar{W}(k)$ for every $k$ in the window and select the subset $J$ with $\card(J)\geq p-q$ that is equal to $\bar{W}(k)$ most often in the $i$-th window. We denote this $J$ as $J(i)$. Then, we select $\left\lbrace 1,\ldots,p\right\rbrace \setminus J(i)$ as the set of sensors under attack in the $i$-th window. This isolation strategy is  stated in Algorithm \ref{alg:1}.\\[1mm]
\textbf{Example 4:} Consider the nonlinear system subject to measurement noise and sensor attacks
	\begin{equation}
	\left\{\begin{split}
	x^{+}=&\left[ \begin{matrix}\label{e61}
	1&0.1\\
	0&1
	\end{matrix}\right]x+\left[ \begin{matrix}
	0.05 \sin (x_{1}+x_{2})\\
	0.1\sin (x_{1}+x_{2})
	\end{matrix}\right], \\
	y=&\left[ \begin{matrix}
	3&3&6&1.2\\
	0.3&0.6&0.9&12
	\end{matrix}\right]^{\top} x+m+a,
	\end{split}\right.
	\end{equation}
with $m_{i}\sim\mathcal{U}(-0.5,0.5)$ for $i\in\left\lbrace 1,2,3,4\right\rbrace $. Using the design method proposed in \cite{Yang2018a}, we have found that circle-criterion observers of the form (\ref{88}) satisfying Definition \ref{def2} exist for each subset $J\subset\left\lbrace 1,2,3,4\right\rbrace $ with $\card(J)\geq 1$ and $p=4$. It follows that, by Assumption \ref{gg2}, $q=1$. We design a circle-criterion observer for each $J\subset\left\lbrace 1,2,3,4\right\rbrace $ with $\card(J)=3$ and each $S\subset\left\lbrace 1,2,3,4\right\rbrace $ with $\card(S)=2$. Therefore, in total, $\binom{4}{3} + \binom{4}{2} = 10$ observers are designed. We obtain their ISS gains by Monte Carlo simulations, initialize the observers at $\hat{x}(0)=x(0)$, select $(x_{1}(0),x_{2}(0))$ from a standard normal distribution, and fix $\epsilon=0$. We let $N=50,100,200$, and follow the evolution of Algorithm \ref{alg:1}\linebreak for $1000$ time-steps. We attack sensor three, i.e., $W=\left\lbrace 3\right\rbrace $, and let $a_{3}\sim\mathcal{U}(-2,2)$. The isolation results are shown in Figures \ref{fig:1}. In this figure, for visualization only, we depict $\tilde{A}_{i}=\emptyset$ (no isolated sensors) by sensor $0$ being isolated in the $i$-th time window.
\begin{algorithm}[t]
	\caption{Attack Isolation.}
	\label{alg:1}
	\begin{algorithmic}
		\State 1. Design an observer satisfying Definition \ref{def2} for each subset $J\subset\left\lbrace 1,\ldots,p\right\rbrace $ with $\card(J)=p-q$ and each subset $S\subset\left\lbrace 1,\ldots,p\right\rbrace $ with $\card(S)=p-2q$.
		\State 2. Initialize the counter variable $n_{J}(i)=0$ for all $J$ with $\card(J)\geq p-q$ and all $i\in\mathbb{Z}_{>0}$.
		\State 3. Compute $\bar{\pi}_{J}$ for each $J$ with $\card(J)=p-q$ as (\ref{gg2}).
		\State 4. For $i\in\mathbb{Z}_{>0}$ and $\forall k\in \left[ \bar{k}^{*}+(i-1)N,\bar{k}^{*}+iN-1\right] $, compute $\pi_{J}(k)$, $\forall J$ with $\card(J)=p-q$, as
		\begin{equation*}
		\pi_{J}(k)=\max_{S\subset J:\card(S)=p-2q}|\hat{x}_{J}(k)-\hat{x}_{S}(k)|.
		\end{equation*}
		\State 5. For all $k\in \left[ \bar{k}^{*}+(i-1)N,\bar{k}^{*}+iN-1\right] $, take the union of all the subsets $J$ such that $\pi_{J}(k)\leq\bar{\pi}_{J}$:
		\begin{equation*}
		\bar{W}(k)=\underset{J\subset\left\lbrace 1,\ldots,p\right\rbrace :\card(J)=p-q,\pi_{J}(k)\leq\bar{\pi}_{J}}{\bigcup} J.
		\end{equation*}
		\State 6. For $k\in \left[ \bar{k}^{*}+(i-1)N,\bar{k}^{*}+iN-1\right] $, if $\bar{W}(k)=J$ for some $J$ with $\card(J)\geq p-q$, then update its corresponding counter variable as $n_{J}(i)=n_{J}(i)+1$.
		\State 7. For all $i\in\mathbb{Z}_{>0}$, select the subset $J$ with $\card(J)\geq$ $p-q$ that is equal to $\bar{W}(k)$ most often, i.e.,
		\begin{equation*}
		J(i)=\underset{J\in\left\lbrace 1,\ldots,p\right\rbrace :\card(J)\geq p-q}{\argmax} n_{J}(i).
		\end{equation*}
		\State 8. For all $i\in\mathbb{Z}_{>0}$, the set of sensors potentially under attack is given by $\tilde{A}(i) = \left\lbrace 1,\ldots,p\right\rbrace \setminus J(i)$.
		\State 9. For all $i\in\mathbb{Z}_{>0}$, return $\tilde{A}(i)$.
	\end{algorithmic}
\end{algorithm}

\section{Conclusion}

Following the idea of sensor redundancy and multi-observer in \cite{Chong2015}, a general estimation scheme has been proposed for a large class of nonlinear plants and observers, which provides robust estimate of the system state when a sufficiently small subset of sensors are corrupted by (potentially unbounded) attack signals and system plant as well as all sensors are affected by bounded noise. We have posed the multi-observer estimation scheme in terms of the existence of a bank of (local and practical) nonlinear observers with ISS (with respect to disturbances and noise) estimation error dynamics. We have proved that the proposed estimator provides ISS-like estimates of the system state with respect to disturbances only and independent of sensor attacks. This scheme has been proposed in \cite{Chong2015}, for linear systems/observers. Here, we have proposed a unifying framework for a much larger class of nonlinear systems/observers and provided the corresponding stability properties that the estimator yields in the nonlinear case. Using the proposed estimator, we have provided an isolation algorithm to pinpoint sensor attacks during finite time windows. Simulations results have been provided to illustrate the performance of our tools.

\bibliographystyle{ieeetrans}        
\bibliography{Observer}           

%

\end{document}